\documentclass[a4paper]{article}
\RequirePackage[english]{babel}
\RequirePackage[latin1]{inputenc}
\RequirePackage[T1]{fontenc}
\RequirePackage{mathrsfs}
\RequirePackage{amsmath}
\RequirePackage{amssymb}
\RequirePackage{amsbsy}
\RequirePackage{bm}
\begin{document}
%%%%%%%%%%%%%%%%%%%%%%%%%%%%%%%%%%%%%%%%%%%%%%%%%%%%%%%%%%%%%%%%%%%%%%%%%%%%%%%%%%%%%%%%%%%%%%%%%%%
\title{\bf{Metric-Torsional Conformal Gravity}}
\author{Luca Fabbri\\ 
\footnotesize DIPTEM Sez. Metodi e Modelli Matematici, Universit\`{a} di Genova,\\
\footnotesize Piazzale Kennedy Pad. D, 16129 Genova, ITALY and \\
\footnotesize INFN \& Dipartimento di Fisica, Universit{\`a} di Bologna,\\
\footnotesize Via Irnerio 46, 40126 Bologna, ITALY}
\date{}
%%%%%%%%%%%%%%%%%%%%%%%%%%%%%%%%%%%%%%%%%%%%%%%%%%%%%%%%%%%%%%%%%%%%%%%%%%%%%%%%%%%%%%%%%%%%%%%%%%%
\maketitle
%%%%%%%%%%%%%%%%%%%%%%%%%%%%%%%%%%%%%%%%%%%%%%%%%%%%%%%%%%%%%%%%%%%%%%%%%%%%%%%%%%%%%%%%%%%%%%%%%%%
\begin{abstract}
When in general geometric backgrounds the metric is accompanied by torsion, the metric conformal properties should correspondingly be followed by analogous torsional conformal properties; however a combined metric torsional conformal structure has never been found which provides a curvature that is both containing metric-torsional degree of freedom and conformally invariant: in this paper we construct such a metric-torsional conformal curvature. We proceed by building the most general action, then deriving the most general system of field equations; we check their consistency by showing that both conservation laws and trace condition are verified. Final considerations and comments are outlined.
\end{abstract}
%%%%%%%%%%%%%%%%%%%%%%%%%%%%%%%%%%%%%%%%%%%%%%%%%%%%%%%%%%%%%%%%%%%%%%%%%%%%%%%%%%%%%%%%%%%%%%%%%%%
\section*{Introduction}
Of all extended theories of gravitation, one that has a special importance is the one displaying conformal invariance: the theoretical reason for which this extension is elegant is that, of all extensions that can be constructed by considering Lagrangians with more curvatures, the special case given by a conformally symmetric lagrangian is unique, as discussed by Weyl; the phenomenological argument for which such an extension is important is that in this generalization the scale symmetry is related to the property of renormalizability, and thus to the problem of quantization, as discussed by Stelle in \cite{s}; the observational fact for which this extension is interesting is that within this generalization the projective structure gives rise to the possibility of describing in terms of background effects the rotation of galaxies, therefore reducing to geometry the problem of dark matter, as discussed by Mannheim and Kazanas in \cite{m-k}. There are also motivations against a theory of gravitation with conformal invariance related to the fact that the universe appears not to possess such a symmetry, and so a mechanism of gravitational spontaneous conformal symmetry breaking must be introduced: if the Ricci scalar has a positive value in vacuum, at least asymptotically, then such a mechanism is possible, as it has been discussed in \cite{e-f-p/1}; this mechanism provides not only conformal but also gauge symmetry breaking, as it is further discussed in \cite{e-f-p/2}. A basic introduction to gravitational conformal theories and their developments is for example in \cite{f} and references therein.

On the other hand however, although theories of gravity are fundamentally metric nevertheless another essential component is torsion, so that not only metric but also torsion has to undergo to conformal transformations: one of the possibilities is therefore to consider that like the metric also torsion has conformal transformations, and this circumstance is called strong conformal transformation; yet another possibility is to think that only the metric is genuinely conformally transforming whereas torsion is left unchanged, and this situation is called weak conformal transformation. A discussion about the relationships between these instances of conformal transformations and further special cases is for example in \cite{sh} and in some of the references therein.

In this scenario, if we want to consider a gravitational conformal model in which both metric and torsion conformal transformations are defined, the central idea is to define a generalized metric-torsional curvature that is conformally covariant: however this simple idea is difficult to be realized because if it is known that, on the one hand, from the Riemann metric curvature tensor we can take the irreducible part which turns out to be conformally invariant, it is also true that, on the other hand, from the Riemann-Cartan metric-torsional curvature tensor we may take the irreducible part which now is not conformally invariant whatsoever. The idea is then to consider the Riemann-Cartan metric-torsional curvature tensor modified in such a way that its irreducible part does turn out to be conformally invariant in $(1+3)$-dimensional spacetimes.

In the present paper we shall find this metric-torsional conformal curvature of the $(1+3)$-dimensional spacetime. Then we shall study the metric-torsional conformal gravitational dynamics by obtaining the field equations, whose consistency is checked in terms of the conservation laws and trace conditions.
%%%%%%%%%%%%%%%%%%%%%%%%%%%%%%%%%%%%%%%%%%%%%%%%%%%%%%%%%%%%%%%%%%%%%%%%%%%%%%%%%%%%%%%%%%%%%%%%%%%
%%%%%%%%%%%%%%%%%%%%%%%%%%%%%%%%%%%%%%%%%%%%%%%%%%%%%%%%%%%%%%%%%%%%%%%%%%%%%%%%%%%%%%%%%%%%%%%%%%%
\section{Metric-Torsional Conformal Curvature Tensor}
In this paper, the Riemann-Cartan metric-torsional geometry is defined in terms of a metric $g_{\alpha\beta}$ and a metric-compatible connection $\Gamma^{\mu}_{\alpha\sigma}$ where metric and metric-compatible connection are taken to be independent: metric-compatibility means that by applying to the metric tensor the covariant derivative associated to the connection the result vanishes; on the other hand, in defining the covariant derivatives, the two lower indices of a connection have two different roles, and thus the connection is not symmetric in these two indices and its antisymmetric part in those two indices is a tensor that does not vanish, known as Cartan torsion tensor $Q_{\sigma\rho\alpha}$ decomposable as
\begin{eqnarray}
&\Gamma^{\sigma}_{\phantom{\sigma}\rho\alpha}=
\frac{1}{2}g^{\sigma\theta}[Q_{\rho\alpha\theta}+Q_{\alpha\rho\theta}+Q_{\theta\rho\alpha}
+(\partial_{\rho}g_{\alpha\theta}+\partial_{\alpha}g_{\rho\theta}-\partial_{\theta}g_{\rho\alpha})]
\label{connection}
\end{eqnarray}
showing that because of Cartan torsion the metric and the metric-compatible connections are independent indeed. An equivalent formalism can be introduced, in which we consider the constant Minkowskian metric $\eta_{ij}$ and a basis of vierbein $e_{\alpha}^{i}$ such that we have the relationship $e_{\alpha}^{p}e_{\nu}^{i}\eta_{pi}=g_{\alpha\nu}$ together with the spin-connection $\omega^{ip}_{\phantom{ip}\alpha}$ and vierbein and spin-connection are again taken to be independent: then we have the correspondent metric-compatibilities for which by applying to the Minkowskian metric and the vierbein the covariant derivative associated to the spin-connection the results vanish; respectively we have the antisymmetry of the spin-connection $\omega^{ip}_{\phantom{ip}\alpha}=-\omega^{pi}_{\phantom{pi}\alpha}$ with spin-connection and metric-compatible connection related by the following formula
\begin{eqnarray}
&\omega^{i}_{\phantom{i}p\alpha}=
e^{i}_{\sigma}(\Gamma^{\sigma}_{\rho\alpha}e^{\rho}_{p}+\partial_{\alpha}e^{\sigma}_{p})
\label{spin-connection}
\end{eqnarray} 
showing that the vierbein and the spin-connection are independent. The former formalism indicated with Latin letters and the latter formalism indicated with Greek letters are respectively denoted as spacetime formalism and world formalism, and they are equivalent; in these equivalent formalisms the independence between metric and connection is equivalent to the independence between vierbein and spin-connection. For a more extensive introduction we refer to \cite{f}.

Now the conformal transformation is given by requiring that the line element is stretched by a given function $\sigma$ and therefore we have that for the metric it is expressed by
\begin{eqnarray}
&g_{\alpha\beta}\rightarrow\sigma^{2}g_{\alpha\beta}
\end{eqnarray}
while by defining $\ln{\sigma}=\phi$ we have that for the torsion tensor it is given by the following
\begin{eqnarray}
&Q^{\sigma}_{\phantom{\sigma}\rho\alpha}\rightarrow Q^{\sigma}_{\phantom{\sigma}\rho\alpha}
+q(\delta^{\sigma}_{\rho}\partial_{\alpha}\phi-\delta^{\sigma}_{\alpha}\partial_{\rho}\phi)
\end{eqnarray}
in terms of the parameter $q$ as the most general possible; henceforth from the relationship (\ref{connection}) it is possible to see what is the conformal transformation for the connection in its most general form. Given that there is no conformal transformation for the constant Minkowskian matrix then the conformal transformation for the vierbein is simply
\begin{eqnarray}
&e_{\alpha}^{k}\rightarrow\sigma e_{\alpha}^{k}
\end{eqnarray}
as it is clear; therefore by employing the relationship (\ref{spin-connection}) it is possible to get the conformal transformation for the spin-connection. For a general discussion about the most general conformal transformation for the metric-torsional or equivalently the vierbein-torsional system we refer the reader to reference \cite{sh}.

In this framework, the Riemann-Cartan metric-torsional curvature tensor is
\begin{eqnarray}
\nonumber
&G^{i}_{\phantom{i}k\mu\nu}=G^{\rho}_{\phantom{\rho}\xi\mu\nu}e^{i}_{\rho}e^{\xi}_{k}=\\
\nonumber
&=(\partial_{\mu}\Gamma^{\rho}_{\xi\nu}-\partial_{\nu}\Gamma^{\rho}_{\xi\mu}
+\Gamma^{\rho}_{\sigma\mu}\Gamma^{\sigma}_{\xi\nu}
-\Gamma^{\rho}_{\sigma\nu}\Gamma^{\sigma}_{\xi\mu})e^{i}_{\rho}e^{\xi}_{k}\equiv\\
&\equiv\partial_{\mu}\omega^{i}_{\phantom{i}k\nu}-\partial_{\nu}\omega^{i}_{\phantom{i}k\mu}
+\omega^{i}_{\phantom{i}a\mu}\omega^{a}_{\phantom{a}k\nu}
-\omega^{i}_{\phantom{i}a\nu}\omega^{a}_{\phantom{a}k\mu}
\label{curvature}
\end{eqnarray}
within which there is the implicit presence of the Cartan torsion tensor; the Riemann-Cartan curvature tensor is antisymmetric in both the first and second couple of indices while Cartan torsion is antisymmetric in the second couple of indices, and accordingly the Riemann-Cartan curvature has one independent contraction that is chosen to be given in the form $G^{\rho}_{\phantom{\rho}\mu\rho\nu}=G^{i}_{\phantom{i}\mu\rho\nu}e^{\rho}_{i}=G_{\mu\nu}$ whose contraction is $G_{\eta\nu}g^{\eta\nu}=G$ while Cartan torsion has one independent contraction chosen to be given by $Q^{\rho}_{\phantom{\rho}\rho\nu}=Q_{\nu}$ setting our convention. The commutator of covariant derivatives can be expressed in terms of both curvature and torsion and thus the cyclic permutations of commutators of commutators of covariant derivatives gives an inner relationship between these two tensors as 
\begin{eqnarray}
\nonumber
&(D_{\mu}G^{\nu}_{\phantom{\nu}\iota \sigma \rho}
-G^{\nu}_{\phantom{\nu}\iota \beta \mu}Q^{\beta}_{\phantom{\beta}\sigma\rho})
+(D_{\sigma}G^{\nu}_{\phantom{\nu}\iota \rho \mu}
-G^{\nu}_{\phantom{\nu}\iota \beta \sigma}Q^{\beta}_{\phantom{\beta}\rho\mu})+\\
&+(D_{\rho}G^{\nu}_{\phantom{\nu}\iota \mu \sigma}
-G^{\nu}_{\phantom{\nu}\iota \beta \rho}Q^{\beta}_{\phantom{\beta}\mu\sigma})\equiv0\\
\nonumber
&(D_{\sigma}Q^{\rho}_{\phantom{\rho}\mu \nu}
+Q^{\rho}_{\phantom{\rho}\sigma \pi}Q^{\pi}_{\phantom{\pi}\mu \nu}
+G^{\rho}_{\phantom{\rho}\sigma\mu\nu})
+(D_{\nu}Q^{\rho}_{\phantom{\rho} \sigma \mu}
+Q^{\rho}_{\phantom{\rho}\nu \pi}Q^{\pi}_{\phantom{\pi}\sigma \mu}
+G^{\rho}_{\phantom{\rho}\nu\sigma\mu})+\\
&+(D_{\mu}Q^{\rho}_{\phantom{\rho} \nu \sigma}
+Q^{\rho}_{\phantom{\rho}\mu \pi}Q^{\pi}_{\phantom{\pi} \nu \sigma}
+G^{\rho}_{\phantom{\rho}\mu\nu\sigma})\equiv0
\end{eqnarray}
known as Jacobi-Bianchi identities, which will be used in the following.

Now if we were to start from the purely metric curvature $R^{\rho}_{\phantom{\rho}\eta\mu\nu}$ and employ its contraction to construct a metric curvature $W^{\rho}_{\phantom{\rho}\eta\mu\nu}$ with the same symmetries but irreducible then we would get a tensor with the property of being conformally covariant, but as torsion is included within the connection the most straightforward generalization of the metric-torsional irreducible curvature would not be conformal invariant any longer; to solve this problem and get a metric-torsional conformal curvature, the simplest idea would be to find a way in which torsion should be included through the connection implicitly as for the metric-torsional curvature $G^{\rho}_{\phantom{\rho}\eta\mu\nu}$ and added explicitly to get a metric-torsional modified curvature $M^{\rho}_{\phantom{\rho}\eta\mu\nu}$ with the same symmetries but such that its irreducible part $T^{\rho}_{\phantom{\rho}\eta\mu\nu}$ would be conformally covariant: this can actually be done by starting from the metric-torsional curvature $G^{\rho}_{\phantom{\rho}\eta\mu\nu}$ plus torsion defining the metric-torsional modified curvature tensor as given by the following form
\begin{eqnarray}
&M_{\alpha\beta\mu\nu}
=G_{\alpha\beta\mu\nu}+(\frac{1-q}{3q})(Q_{\beta}Q_{\alpha\mu\nu}-Q_{\alpha}Q_{\beta\mu\nu})
\label{modifiedcurvature}
\end{eqnarray}
antisymmetric in both the first and second couple of indices, whose contractions are given by $M^{\rho}_{\phantom{\rho}\mu\rho\nu}=M_{\mu\nu}$ with contraction $M_{\eta\nu}g^{\eta\nu}=M$ and from which we construct the metric-torsional irreducible curvature tensor as given by
\begin{eqnarray}
&T_{\alpha\beta\mu\nu}=M_{\alpha\beta\mu\nu}
-\frac{1}{2}(M_{\alpha[\mu}g_{\nu]\beta}-M_{\beta[\mu}g_{\nu]\alpha})
+\frac{1}{12}M(g_{\alpha[\mu}g_{\nu]\beta}-g_{\beta[\mu}g_{\nu]\alpha})
\label{conformalcurvature}
\end{eqnarray}
antisymmetric in both the first and second couple of indices and irreducible, and also conformally covariant as desired. In this way we have constructed the metric-torsional conformal curvature in $(1+3)$-dimensional spacetimes.

Notice that there are two special cases of the parameter $q$ that have to be considered: the first is given by the fact that for $q=0$ it is not possible to define a metric-torsional modified curvature and therefore the metric-torsional conformal curvature, at least in this most straightforward manner, telling that torsion cannot have the simplest weak conformal transformations, but instead torsion must have the most general strong conformal transformation, if we want to have any chance to construct, at least in the simplest case, a metric-torsional conformal curvature; the second is given by $q=1$ where the metric-torsional curvature does not need to be modified in order for its irreducible part to be conformally invariant, telling that the torsional strong conformal transformation is actually a conformal transformation, because the metric and torsional conformal transformations are able to compensate one another. A final comment is that in the limit for vanishing torsion the present metric-torsional conformal curvature reduces to the purely metric conformal curvature as expressed by Weyl tensor.
%%%%%%%%%%%%%%%%%%%%%%%%%%%%%%%%%%%%%%%%%%%%%%%%%%%%%%%%%%%%%%%%%%%%%%%%%%%%%%%%%%%%%%%%%%%%%%%%%%%
%%%%%%%%%%%%%%%%%%%%%%%%%%%%%%%%%%%%%%%%%%%%%%%%%%%%%%%%%%%%%%%%%%%%%%%%%%%%%%%%%%%%%%%%%%%%%%%%%%%
\section{Metric-Torsional Conformal Gravity;\\ Energy-Spin Conformal Dynamical Systems}
Up to now we have been able to obtain the quantity that describes the metric-torsional conformal gravitational background, and in the following we shall study the case in which this background contains energy-spin matter with conformal dynamics which for the moment will be taken in its most general form.

Since (\ref{conformalcurvature}) is a conformal tensor then we shall employ it to work out all possible invariants: because it is irreducible then we will have to take the product of two of them contracting indices of one another; and because of its anti-symmetries in both the first and second couple of indices then we will have that the independent invariants are $T^{\alpha\beta\mu\nu}T_{\alpha\beta\mu\nu}$, $T^{\alpha\beta\mu\nu}T_{\mu\nu\alpha\beta}$, $T^{\alpha\beta\mu\nu}T_{\alpha\mu\beta\nu}$ so that it will be in terms of the parameters $A$, $B$, $C$ that the most general invariant is given by $AT^{\alpha\beta\mu\nu}T_{\alpha\beta\mu\nu}
+BT^{\alpha\beta\mu\nu}T_{\mu\nu\alpha\beta}+CT^{\alpha\beta\mu\nu}T_{\alpha\mu\beta\nu}$ as a straightforward analysis may show. Therefore it is useful to define the parametric quantity
\begin{eqnarray}
&P_{\alpha\beta\mu\nu}=AT_{\alpha\beta\mu\nu}+BT_{\mu\nu\alpha\beta}+\frac{C}{4}(T_{\alpha\mu\beta\nu}-T_{\beta\mu\alpha\nu}+T_{\beta\nu\alpha\mu}-T_{\alpha\nu\beta\mu})
\label{parametricconformalcurvature}
\end{eqnarray}
in terms of the parameters $A$, $B$, $C$, antisymmetric in both the first and second couple of indices, irreducible and conformally covariant: in terms of this parametric conformal tensor $P_{\alpha\beta\mu\nu}$ the most general invariant we wrote above reduces to the form given by $T^{\alpha\beta\mu\nu}P_{\alpha\beta\mu\nu}$ and so the most general action is
\begin{eqnarray}
&S=\int[kT^{\alpha\beta\mu\nu}P_{\alpha\beta\mu\nu}+L_{\mathrm{matter}}]\sqrt{|g|}dV
\label{action}
\end{eqnarray}
with constant $k$ and complemented by the material lagrangian, and where it is over the volume of the spacetime that the integral is taken. By varying this action with respect to metric and connection or equivalently vierbein and spin-connection one obtains the field equations
\begin{eqnarray}
\nonumber
&2k[P^{\theta\sigma\rho\alpha}T_{\theta\sigma\rho}^{\phantom{\theta\sigma\rho}\mu}
-\frac{1}{4}g^{\alpha\mu}P^{\theta\sigma\rho\beta}T_{\theta\sigma\rho\beta}
+P^{\mu\sigma\alpha\rho}M_{\sigma\rho}+\\
\nonumber
&+(\frac{1-q}{3q})
(D_{\nu}(2P^{\mu\rho\alpha\nu}Q_{\rho}
-g^{\mu\alpha}P^{\nu\theta\rho\sigma}Q_{\theta\rho\sigma}
+g^{\mu\nu}P^{\alpha\theta\rho\sigma}Q_{\theta\rho\sigma})+\\
&+Q_{\nu}(2P^{\mu\rho\alpha\nu}Q_{\rho}
-g^{\mu\alpha}P^{\nu\theta\rho\sigma}Q_{\theta\rho\sigma}
-P^{\mu\nu\rho\sigma}Q^{\alpha}_{\phantom{\alpha}\rho\sigma}))]=\frac{1}{2}T^{\alpha\mu}
\label{energy}\\
\nonumber
&4k[D_{\rho}P^{\alpha\beta\mu\rho}+Q_{\rho}P^{\alpha\beta\mu\rho}
-\frac{1}{2}Q^{\mu}_{\phantom{\mu}\rho\theta}P^{\alpha\beta\rho\theta}-\\
&-(\frac{1-q}{3q})(Q_{\rho}P^{\rho[\alpha\beta]\mu}
-\frac{1}{2}Q_{\sigma\rho\theta}g^{\mu[\alpha}P^{\beta]\sigma\rho\theta})]=S^{\mu\alpha\beta}
\label{spin}
\end{eqnarray}
in terms of the parameter $q$ and the constant $k$ and where $T^{\mu\nu}$ and $S^{\rho\mu\nu}$ are the energy and spin densities of the matter conformal field; as Einstein equations are generalized by Sciama-Kibble equations, similarly Weyl equations are generalized by this system of equations, with the difference that Einstein equations tell how energy is the source of curvature while Sciama-Kibble equations tell how spin is the source of torsion whereas Weyl equations tell how energy is the source of both curvature and torsion while the new system of equations tell how spin is the source of both curvature and torsion. As far as we are concerned, this seems a genuine property of matter fields in conformal dynamical systems.

Finally by taking into account the Jacobi-Bianchi identities in their fully contracted form we have that the field equations (\ref{energy}-\ref{spin}) are converted into conservation laws that are given in the following form
\begin{eqnarray}
&D_{\mu}T^{\mu\rho}+Q_{\mu}T^{\mu\rho}-T_{\mu\sigma}Q^{\sigma\mu\rho}
+S_{\beta\mu\sigma}G^{\sigma\mu\beta\rho}=0
\label{conservationlawenergy}\\
&D_{\rho}S^{\rho\mu\nu}+Q_{\rho}S^{\rho\mu\nu}
+\frac{1}{2}T^{[\mu\nu]}=0
\label{conservationlawspin}
\end{eqnarray}
with trace condition as another conservation law
\begin{eqnarray}
&(1-q)(D_{\mu}S_{\nu}^{\phantom{\nu}\nu\mu}+Q_{\mu}S_{\nu}^{\phantom{\nu}\nu\mu})
+\frac{1}{2}T_{\mu}^{\phantom{\mu}\mu}=0
\label{trace}
\end{eqnarray}
and the whole set of conservation laws have to be satisfied once the matter conformal field equations are eventually given: it is important to notice that the general conservation laws (\ref{conservationlawenergy}-\ref{conservationlawspin}) are now accompanied by an additional conservation law for the trace (\ref{trace}) because general coordinate transformations are now accompanied by general conformal transformations, and the lagrangian formalism tells that for any symmetry a conservation law follows; it is also important to remark that as the energy is not symmetric because its antisymmetric part is related to the spin tensor through (\ref{conservationlawspin}) analogously the energy is not traceless because its trace is related to the spin trace vector through (\ref{trace}), and because the constrain constituted by the trace condition (\ref{trace}) is a conservation law then this constraint is dynamically implemented within the model. In following papers we shall consider specific matter fields to better 
discuss these issues.
%%%%%%%%%%%%%%%%%%%%%%%%%%%%%%%%%%%%%%%%%%%%%%%%%%%%%%%%%%%%%%%%%%%%%%%%%%%%%%%%%%%%%%%%%%%%%%%%%%%
%%%%%%%%%%%%%%%%%%%%%%%%%%%%%%%%%%%%%%%%%%%%%%%%%%%%%%%%%%%%%%%%%%%%%%%%%%%%%%%%%%%%%%%%%%%%%%%%%%%
\section*{Conclusion}
In this paper we have been able to construct the metric-torsional conformal curvature of the $(1+3)$-dimensional spacetime; we have studied the corresponding metric-torsional conformal gravity by obtaining the field equations in presence of energy-spin matter with conformal dynamics, then we have performed a consistency check in terms of both conservation laws and trace condition and a discussion about these conservation laws has been sketched.
%%%%%%%%%%%%%%%%%%%%%%%%%%%%%%%%%%%%%%%%%%%%%%%%%%%%%%%%%%%%%%%%%%%%%%%%%%%%%%%%%%%%%%%%%%%%%%%%%%%
%%%%%%%%%%%%%%%%%%%%%%%%%%%%%%%%%%%%%%%%%%%%%%%%%%%%%%%%%%%%%%%%%%%%%%%%%%%%%%%%%%%%%%%%%%%%%%%%%%%

%%%%%%%%%%%%%%%%%%%%%%%%%%%%%%%%%%%%%%%%%%%%%%%%%%%%%%%%%%%%%%%%%%%%%%%%%%%%%%%%%%%%%%%%%%%%%%%%%%%
\end{document}